\documentclass[twocolumn]{aastex631}

\usepackage{longtable}
\usepackage[flushleft]{threeparttable}
\usepackage{tabularx}
\usepackage{multirow}
\usepackage{graphicx}
\usepackage{amsmath,amssymb}
\usepackage{xcolor}
\usepackage{units}
\usepackage{epstopdf}
\usepackage{hyperref}
\usepackage{multirow}
\usepackage{url}
\usepackage{subfigure}
\usepackage{rotating}
\usepackage{gensymb}
\usepackage{enumitem}\setlist[description]{font=\textendash\enskip\scshape\bfseries}

\definecolor{navyblue}{rgb}{0.0, 0.0, 0.5}

\begin{document}

\title{Optimizing Cadences with Realistic Light Curve Filtering for Serendipitous Kilonova Discovery with Vera Rubin Observatory}

\author[0000-0003-3768-7515]{Igor Andreoni}
\affil{Division of Physics, Mathematics and Astronomy, California Institute of Technology, Pasadena, CA 91125, USA}
\email{andreoni@caltech.edu}

\author[0000-0002-8262-2924]{Michael W. Coughlin}
\affil{School of Physics and Astronomy, University of Minnesota, Minneapolis, Minnesota 55455, USA}

\author[0000-0002-4694-7123]{Mouza Almualla}
\affil{Department of Physics, American University of Sharjah, PO Box 26666, Sharjah, UAE}

\author[0000-0001-8018-5348]{Eric C. Bellm}
\affiliation{DIRAC Institute, Department of Astronomy, University of Washington, 3910 15th Avenue NE, Seattle, WA 98195, USA} 

\author{Federica B. Bianco}
\affiliation{Department of Physics and Astronomy, University of Delaware, Newark, DE, 19716, USA}
\affiliation{Joseph R. Biden, Jr. School of Public Policy and Administration, University of Delaware, Newark, DE, 19716, USA}
\affiliation{Data Science Institute, University of Delaware, Newark, DE, 19716, USA}

\author{Mattia Bulla}
\affiliation{The Oskar Klein Centre, Department of Astronomy, Stockholm University, AlbaNova, SE-106 91 Stockholm, Sweden}

\author{Antonino Cucchiara}
\affil{College of Marin, 120 Kent Avenue, Kentfield 94904 CA, USA}

\author[0000-0003-2374-307X]{Tim Dietrich}
\affiliation{Institut f\"{u}r Physik und Astronomie, Universit\"{a}t Potsdam, 14476 Potsdam, Germany}
\affiliation{Max Planck Institute for Gravitational Physics (Albert Einstein Institute), Am M\"uhlenberg 1, Potsdam 14476, Germany}

\author{Ariel Goobar}
\affil{The Oskar Klein Centre, Department of Physics, Stockholm University, AlbaNova, SE-106 91 Stockholm, Sweden}

\author[0000-0002-7252-3877]{Erik C. Kool}
\affiliation{The Oskar Klein Centre, Department of Astronomy, Stockholm University, AlbaNova, SE-106 91 Stockholm, Sweden}

\author[0000-0002-0514-5650]{Xiaolong Li}
\affiliation{Department of Physics and Astronomy, University of Delaware, Newark, DE, 19716, USA}

\author{Fabio Ragosta}
\affiliation{INAF-Osservatorio Astronomico di Capodimonte, via Moiariello 16, I-80131, Naples, Italy}

\author[0000-0002-3498-2167]{Ana Sagu{\'e}s-Carracedo}
\affiliation{The Oskar Klein Centre, Department of Physics, Stockholm University, AlbaNova, SE-106 91 Stockholm, Sweden}

\author[0000-0001-9898-5597]{Leo P. Singer}
\affiliation{Astrophysics Science Division, NASA Goddard Space Flight Center, MC 661, Greenbelt, MD 20771, USA}
\affiliation{Joint Space-Science Institute, University of Maryland, College Park, MD 20742, USA}

\begin{abstract}
Current and future optical and near-infrared wide-field surveys have the potential of finding kilonovae, the optical and infrared counterparts to neutron star mergers, independently of gravitational-wave or high-energy gamma-ray burst triggers.
The ability to discover fast and faint transients such as kilonovae largely depends on the area observed, the depth of those observations, the number of re-visits per field in a given time frame, and the filters adopted by the survey; it also depends on the ability to perform rapid follow-up observations to confirm the nature of the transients. In this work, we assess kilonova detectability in existing simulations of the LSST strategy for the Vera C. Rubin Wide Fast Deep survey, with focus on comparing rolling to baseline cadences.  
Although currently available cadences can enable the detection of $>300$ kilonovae out to $\sim 1400$\,Mpc over the ten-year survey, we can expect only 3--32 kilonovae similar to GW170817 to be recognizable as fast-evolving transients.  We also explore the detectability of kilonovae over the plausible parameter space, focusing on viewing angle and ejecta masses. We find that observations in redder $izy$ bands are crucial for identification of nearby (within 300\,Mpc) kilonovae that could be spectroscopically classified more easily than more distant sources. Rubin's potential for serendipitous kilonova discovery could be
increased by gain of efficiency with the employment of individual 30\,s exposures (as opposed to $2\times$15\,s snap pairs), with the addition of red-band observations coupled with same-night observations in $g$- or $r$-bands, and possibly with further development of a new rolling-cadence strategy.
\end{abstract}

\section{Introduction}

Binary neutron star (BNS) and neutron star--black hole (NSBH) mergers have long been predicted to be associated with short gamma-ray bursts \citep[GRBs; e.g.,][]{Blinnikov1984}, and  optical/infrared transients called kilonovae \citep[e.g.,][]{Li1998}. Along with claimed evidence for kilonovae in some short-gamma ray bursts \citep[e.g.,][]{Tanvir:2013pia,BeWo2013}, the discovery of a kilonova \citep[e.g.,][]{CoFo2017} associated with the first BNS merger detected in gravitational waves, GW170817 \citep{PhysRevLett.119.161101}, nicely confirmed these predictions. This multi-messenger source marked a watershed moment in astrophysics, with prospects to strongly constrain both the neutron star equation of state \citep[e.g.,][]{Metzger:2017wot, Radice:2017lry, Annala:2017llu, Most:2018hfd, Radice:2018ozg, Tews:2018chv, Essick:2019ldf, Capano:2019eae, DiCo2020} and the Hubble Constant \citep[e.g.,][]{Abbott:2017xzu,Fishbach:2018gjp,HoNa2018,2020ApJ...888...67D,Coughlin:2019vtv,DiCo2020}, amongst many other science cases \citep{Baker:2017hug,Ezquiaga:2017ekz}.

Dynamical ejecta \citep[e.g.,][]{Hotokezaka:2012ze,BaGo2013,Dietrich:2016fpt}, which arise from tidal stripping of the neutron star(s) and the neutron stars contact interface, and post-merger ejecta \citep[e.g.,][]{MePi2008,Fernandez:2014cna,Siegel:2017jug,Fernandez:2018kax}, which arise from accretion disk winds surrounding the remnant object, are characterized by low electron fractions. This scenario favors the production of heavy elements such as lanthanides and actinides via rapid neutron capture (known as the $r$-process), and the decay of these unstable nuclei powers the optical/infrared kilonova \citep[e.g.,][]{LaSc1974,Kasen:2013xka,Barnes:2013wka,Barnes:2016umi,Kasen:2017sxr}.

Questions about the sources of heavy element production in the Universe and diversity in the ejecta of the kilonova population can only be answered by the detection and characterization of a large sample of sources. Unveiling such a population is difficult because kilonovae are rare ($<$ 1\,\% of the core collapse supernova rate), fast (fading $\gtrsim$\,0.5\,mag per day in the optical), and faint transients ($M \gtrsim -16$ at peak), and hence are particularly hard to discover. 
Signatures of kilonovae are mostly found during the follow-up of short GRBs \citep[e.g.,][]{Perley2009} and the follow-up of LIGO/Virgo candidates, although only for GW170817 has a counterpart been identified so far.
Rates of BNS mergers are still highly uncertain, with $R=80-810$\,Gpc$^{-3}$\,yr$^{-1}$ based on GW observations \citep{LIGO_GWTC2_pop_2020arXiv}; empirical limits on kilonovae rates by optical surveys are nearing the upper end of the gravitational-wave measurements \citep{AnKo2020, AnCo2021arXiv}.

The advent of Vera C. Rubin Observatory \citep{2019ApJ...873..111I} presents us with a great opportunity to identify a population of kilonovae independent of any gravitational-wave or gamma-ray burst trigger, thanks to the unprecedented volume that the Legacy Survey of Space and Time (LSST) will be able to probe \citep[see e.g.,][]{Andreoni2019PASP, 2019MNRAS.485.4260S}.
Unfortunately, due to their fast fading and intrinsically underluminous properties, ``detection'' is not enough; it is imperative that kilonova candidates found by Rubin Observatory are recognized as such in real time so that follow-up observations can confirm their nature.

Projects exist that are dedicated to fast transient discovery in current wide-field surveys such as the Zwicky Transient Facility \citep[ZTF;][]{Bellm2019PASP, Graham2019PASP}. The ``ZTF Realtime Search and Triggering'' \citep[``\texttt{ZTFReST}''\footnote{\url{github.com/growth-astro/ztfrest}},][]{AnCo2021arXiv} project, for example, employs i) alert queries, ii) forced point-spread-function photometry, and iii) nightly light curve stacking in flux space to discover fast-evolving transients such as kilonovae. 
\texttt{ZTFReST} is proving to be very effective at identifying extragalactic fast transients, having already revealed seven serendipitously-discovered GRB afterglows and at least two supernova shock breakouts in 2020 and in the first three months of 2021 \citep{AnCo2021arXiv}.

In this work, we used the most recent \texttt{OpSim} simulations and a set of new metrics to assess the effectiveness of cadence options for un-triggered, or ``serendipitous,'' kilonova discovery. We employed metrics that both assess Rubin Observatory's ability to simply detect the transients, as well as metrics that are designed to identify a transient as ``fast'' based on its flux evolution. We argue that the latter is the most appropriate metric for potentially maximizing the science output from these rare objects, and we provide suggested cadences based on this metric.

\section{Methods}
\label{sec:methods}

\begin{figure*}[t]
\begin{center}
 \includegraphics[width=0.49\textwidth]{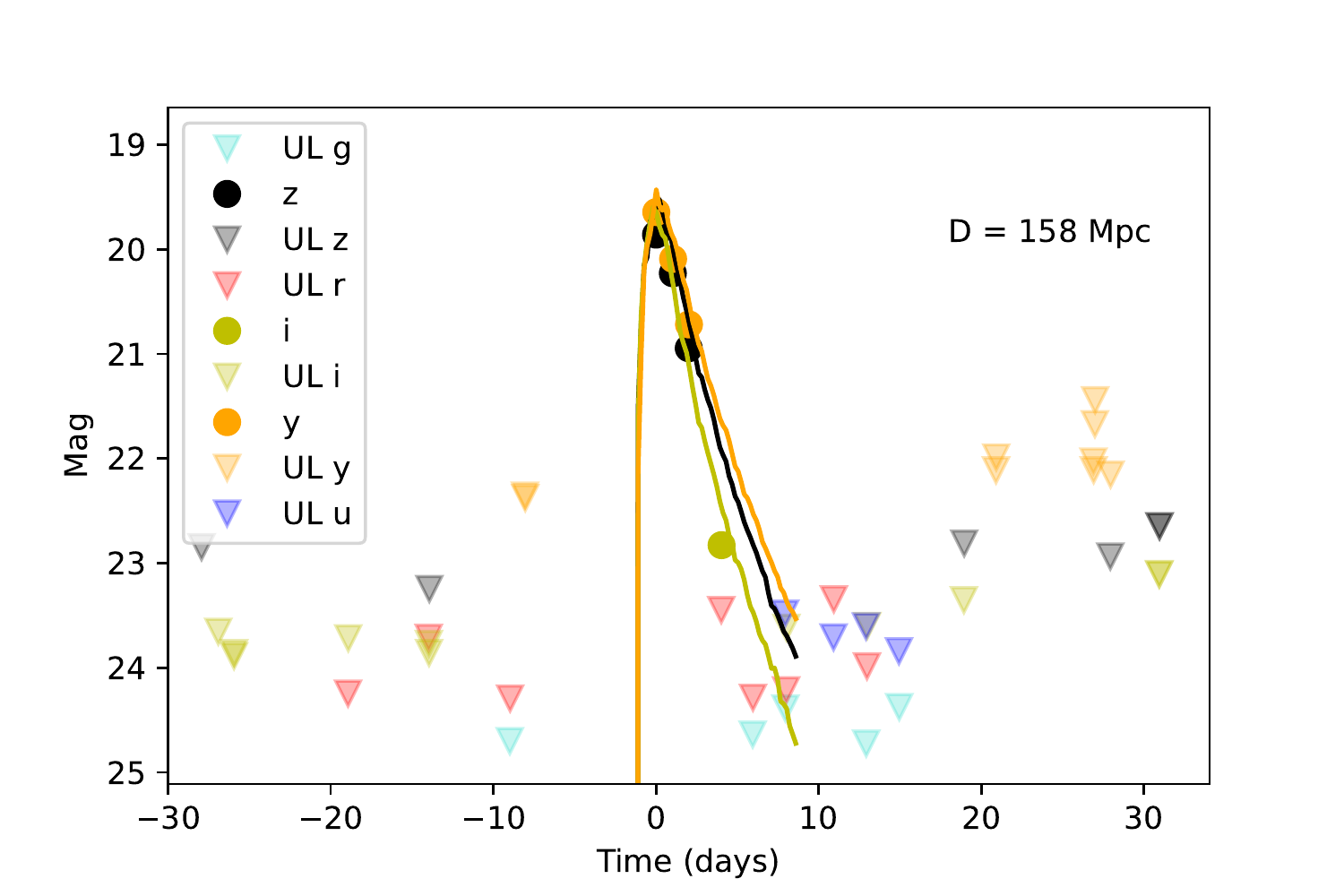}
 \includegraphics[width=0.49\textwidth]{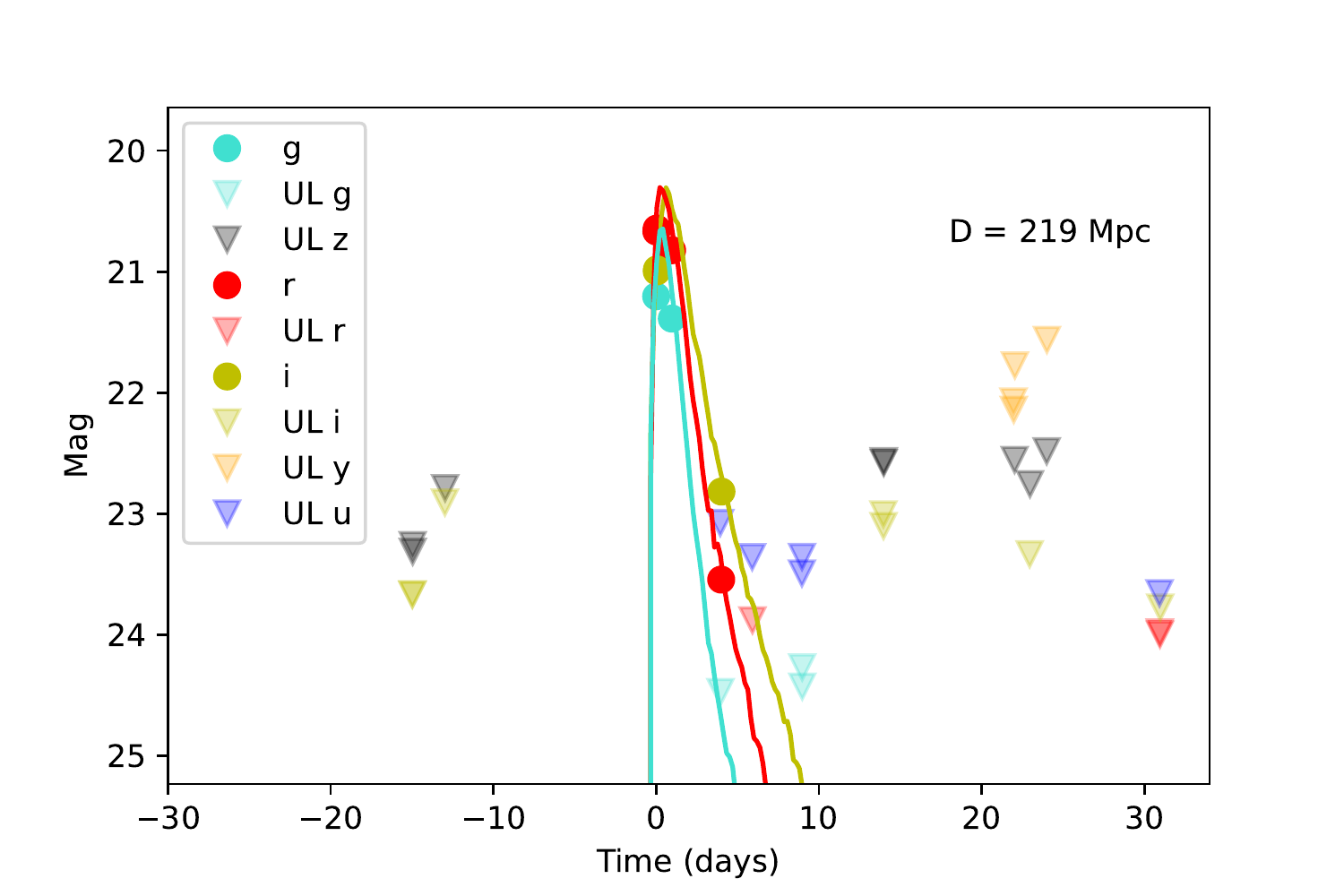}

 \includegraphics[width=0.49\textwidth]{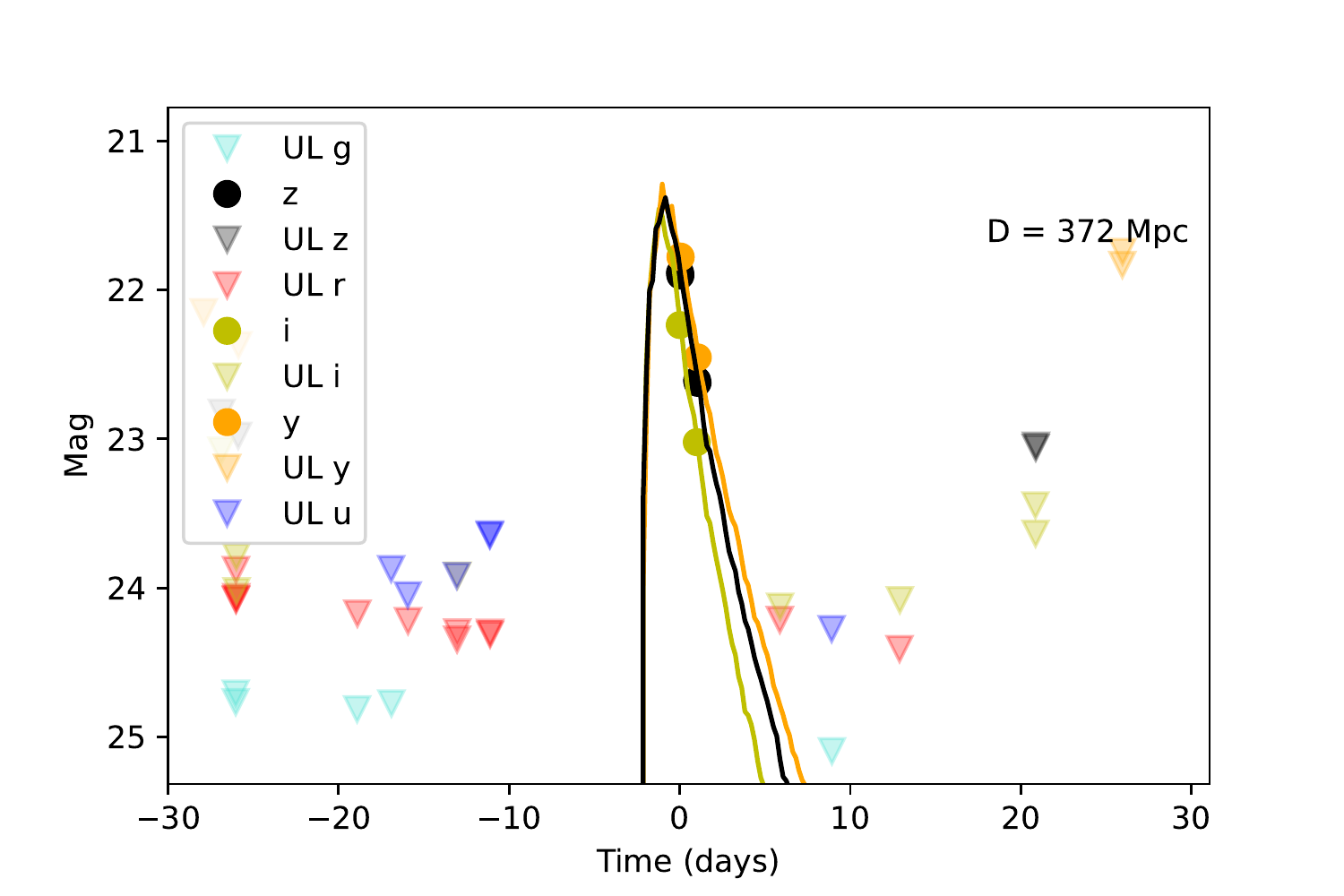}
 \includegraphics[width=0.49\textwidth]{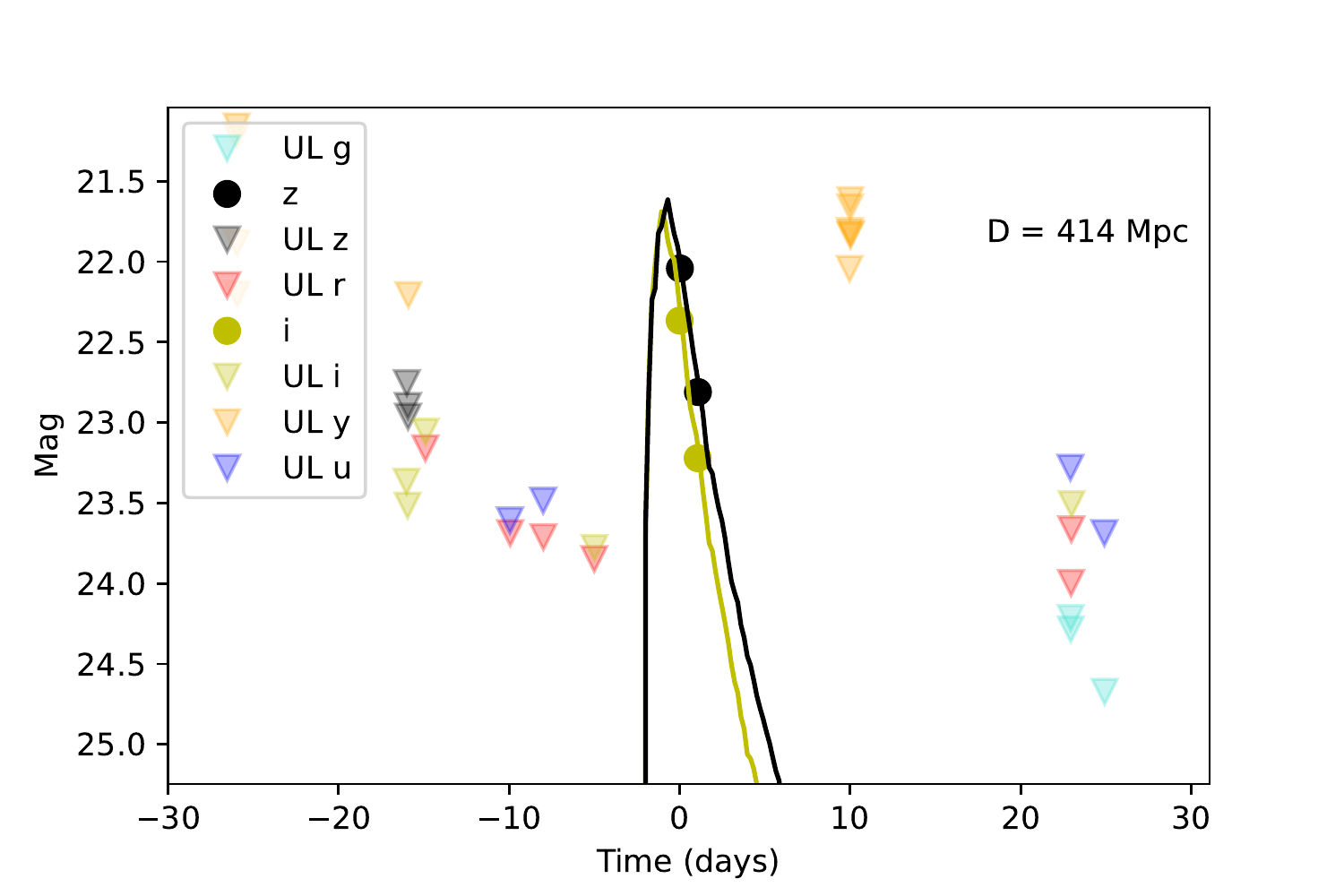}

 \includegraphics[width=0.49\textwidth]{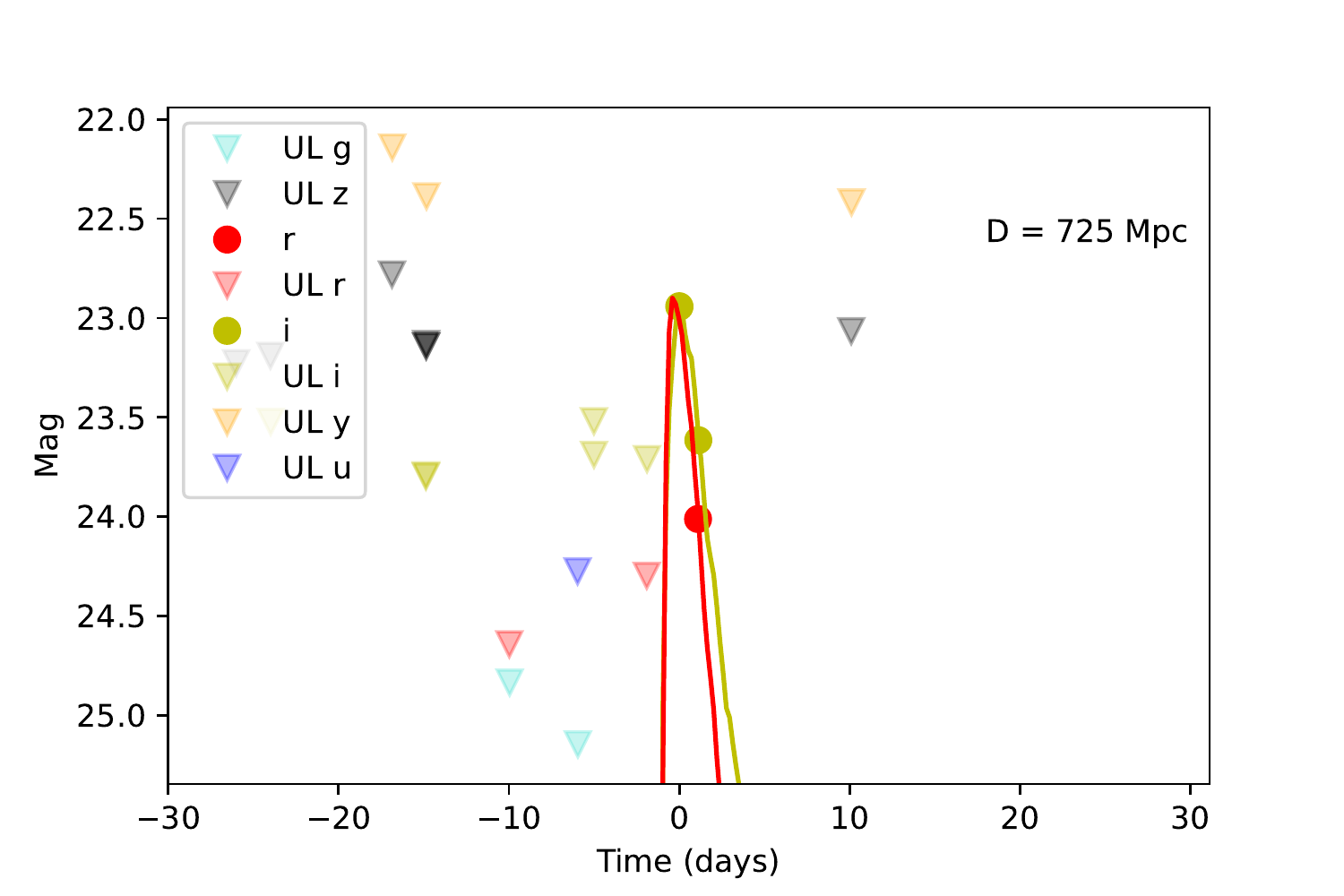}
  \includegraphics[width=0.49\textwidth]{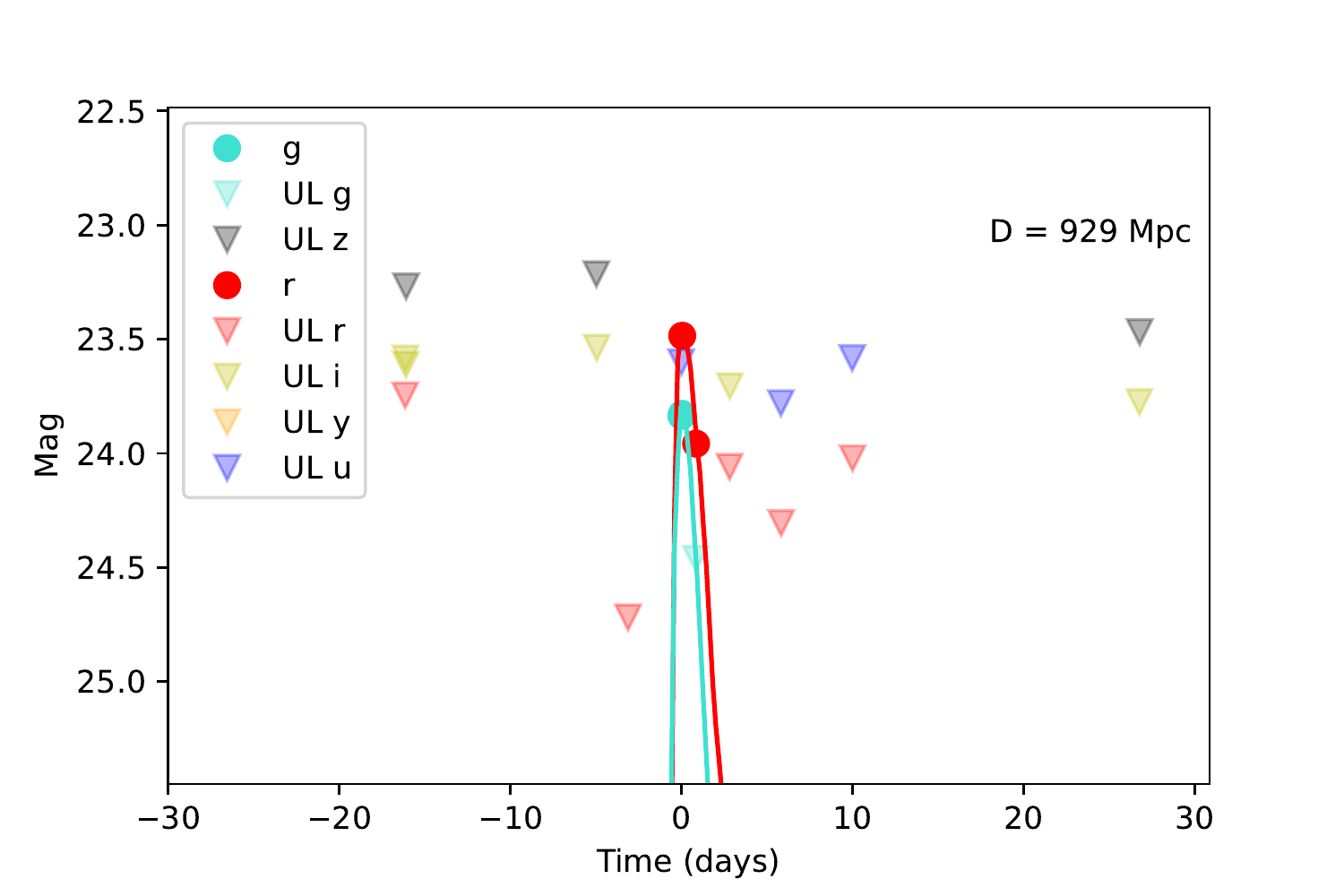}
  
  \caption{Examples of GW170817-like kilonova light curves injected in the baseline cadence (v1.7, individual exposures). Observations from 30 days before peak to 30 days after peak are presented. The light curves were uniformly distributed in volume and uniformly distributed in time throughout the 10-year survey. Circles indicate the detections, solid lines show the simulated light curves in bands where at least one detection is present, and triangles indicate $5\sigma$ upper limits. The luminosity distance at which each light curve is places is also indicated.
  }
 \label{fig:lc_baseline}
\end{center}
\end{figure*}

We used the new \texttt{kneMetrics}\footnote{\url{https://github.com/LSST-nonproject/sims_maf_contrib/blob/master/mafContrib/kneMetrics.py}} to recover synthetic kilonova light curves injected in \texttt{OpSim} simulations. 
The synthetic light curves are taken from \cite{DiCo2020}, which rely on the radiative transfer code \texttt{POSSIS} \citep{Bulla:2019muo}, which vary four parameters over physically viable priors: the dynamic ejecta mass $M_{\rm dyn}$, the disk wind ejecta mass $M_{\rm wind}$, 
the half opening angle of the lanthanide-rich dynamical-ejecta component $\phi$ and the viewing angle $\iota$ \citep[see][for more details about the adopted geometry]{DiCo2020}.
Examples of synthetic light curves injected in the Rubin baseline cadence can be found in Fig.\,\ref{fig:lc_baseline}.

\subsection{Metrics}
\label{subsec:metrics}

To assess kilonova detectability in different cadence simulations, we employed a number of metrics. We improved the existing \texttt{TDEsPopMetric}, designed to inject and recover diverse populations of tidal disruption event light curves, by adding the possibility to inject synthetic transients distributed uniformly in volume (with numbers increasing as a function of distance to the third power), rather than placed at a fixed distance. Light curves at a larger distance share the same properties of those at lower distances, but their apparent luminosity is fainter, making them detectable for shorter times and only when the images' magnitude limits are particularly deep. 

The metrics most relevant to this work, all included as functions in the \texttt{kneMetrics} code, are:

\begin{itemize}

\item \texttt{multi\_detect}: $\geq2$ detections $>5\sigma$
\item \texttt{ztfrest\_simple}: metric that reproduces a discovery algorithm similar\footnote{In \texttt{ZTFReST}, linear fitting is performed, while the \texttt{ztfrest\_simple} metric relies in a more simplistic estimate of the rising or fading rates based on the time and magnitude differences between the brightest and the faintest detected ($>5\sigma$) points in the light curves.} to \texttt{ZTFReST}, in which sources found to be rising faster than 1\,mag\,day$^{-1}$ and fading faster than 0.3\,mag\,day$^{-1}$ are selected
\item \texttt{ztfrest\_simple\_red}: same as \texttt{ztfrest\_simple}, but applied only to $izy$ bands
\item \texttt{ztfrest\_simple\_blue}: same as \texttt{ztfrest\_simple}, but applied only to $ugr$ bands
\end{itemize}

The metrics were deliberately designed to range from standard transient detection (with $\geq 2$ detections) 
which typically provides only spatial information on the celestial coordinates of a source, to methods more likely to lead to source characterization -- in other words, kilonova discovery. Simple detection can be crucial during gravitational-wave follow-up, but is of little use during fast transient searches in the regular survey, especially for transients at large distances. Importantly, the conclusions of this study can be applied to a range of fast transients, including, for example, GRB afterglows and fast blue optical transients (FBOTS), for which light curve sampling with spacing between one hour and one day is crucial.

\begin{figure*}[t]
\begin{center}
 \includegraphics[width=\textwidth]{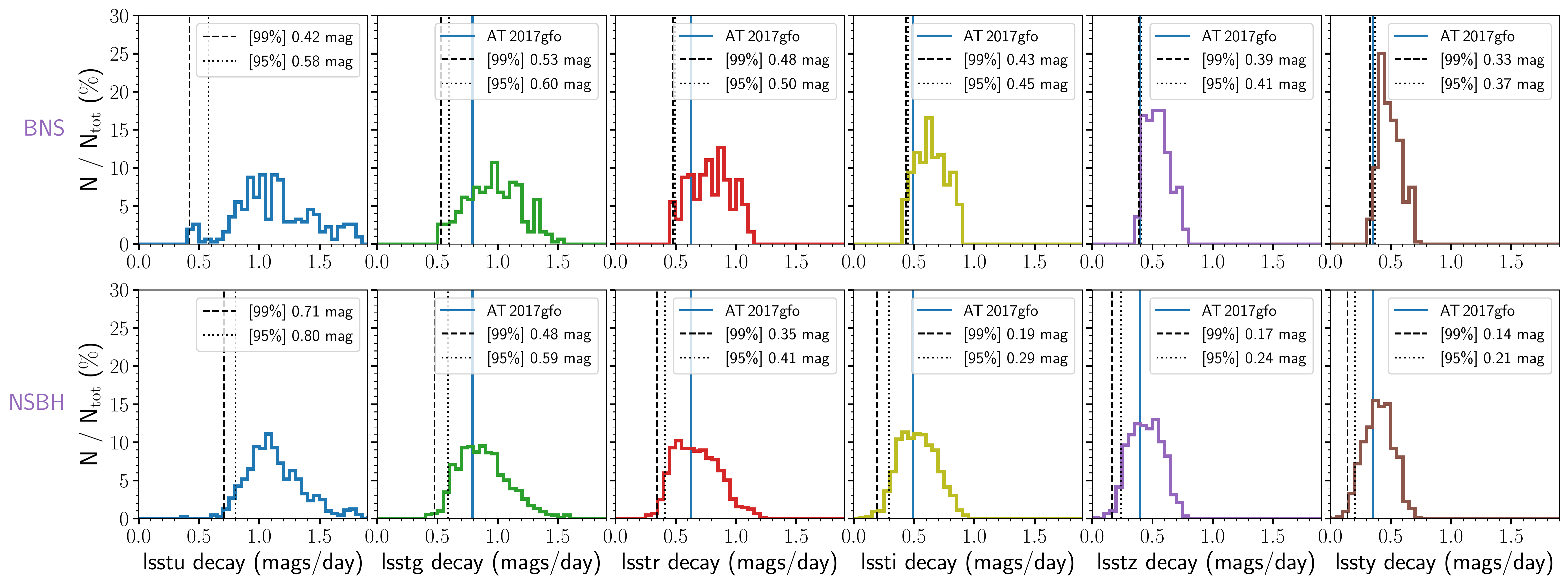}
  \caption{
  A distribution of fade rates was calculated using the kilonova model grid obtained by \cite{Bulla2019}, tailored for BNS (upper panels) and NSBH mergers (lower panels). Averaged decay rates from peak to six days later are shown in Rubin $ugrizy$ bands from left to right.
  Dashed and dot-dashed vertical lines indicate the lower 99\% and 95\% decay rate for each distribution. Blue vertical lines indicate the decay rates for the GW170817 kilonova in each band. A fading-rate threshold of 0.3\,mag\,day$^{-1}$ can enable the identification of kilonovae from BNS mergers ($>99\%$ of the distribution) in all filters.}
 \label{fig:threshold}
\end{center}
\end{figure*}

There are a variety of methods in the literature to promptly identify fast-transient candidates.
For example, methods are being developed for early transient classification via machine learning techniques \citep[e.g.,][]{Muthukrishna:2019wgc}, or as part of the Photometric LSST Astronomical Time Series Classification Challenge \citep[PLAsTiCC;][]{Kessler2019PASP}. Prior work on detecting and identifying fast transients in Rubin LSST \texttt{Opsims} include \cite{Bianco2019}.
A simple but effective strategy to identify transients with rapidly fading or rising light curves can be based on magnitude rise and decay rate measurements. In this work, we consider significant fading rates to be those faster than $0.3$\,mag\,day$^{-1}$, which is the threshold used in real time by the \texttt{ZTFReST} pipeline and is expected to be particularly suitable for the discovery of kilonovae from BNS mergers (Fig.\,\ref{fig:threshold}), or rising rates faster than 1\,mag\,day$^{-1}$, which can separate rapidly evolving transients from most supernovae.  Within \texttt{ZTFReST}, this has greatly helped to separate fast transient candidates from slower, ``contaminant'' sources, with $\sim 30\%$ purity in ``archival'' data searches when considering only fade rates and thresholds tailored for each band. 

\begin{figure*}[t]
\begin{center}
 \includegraphics[width=0.49\textwidth]{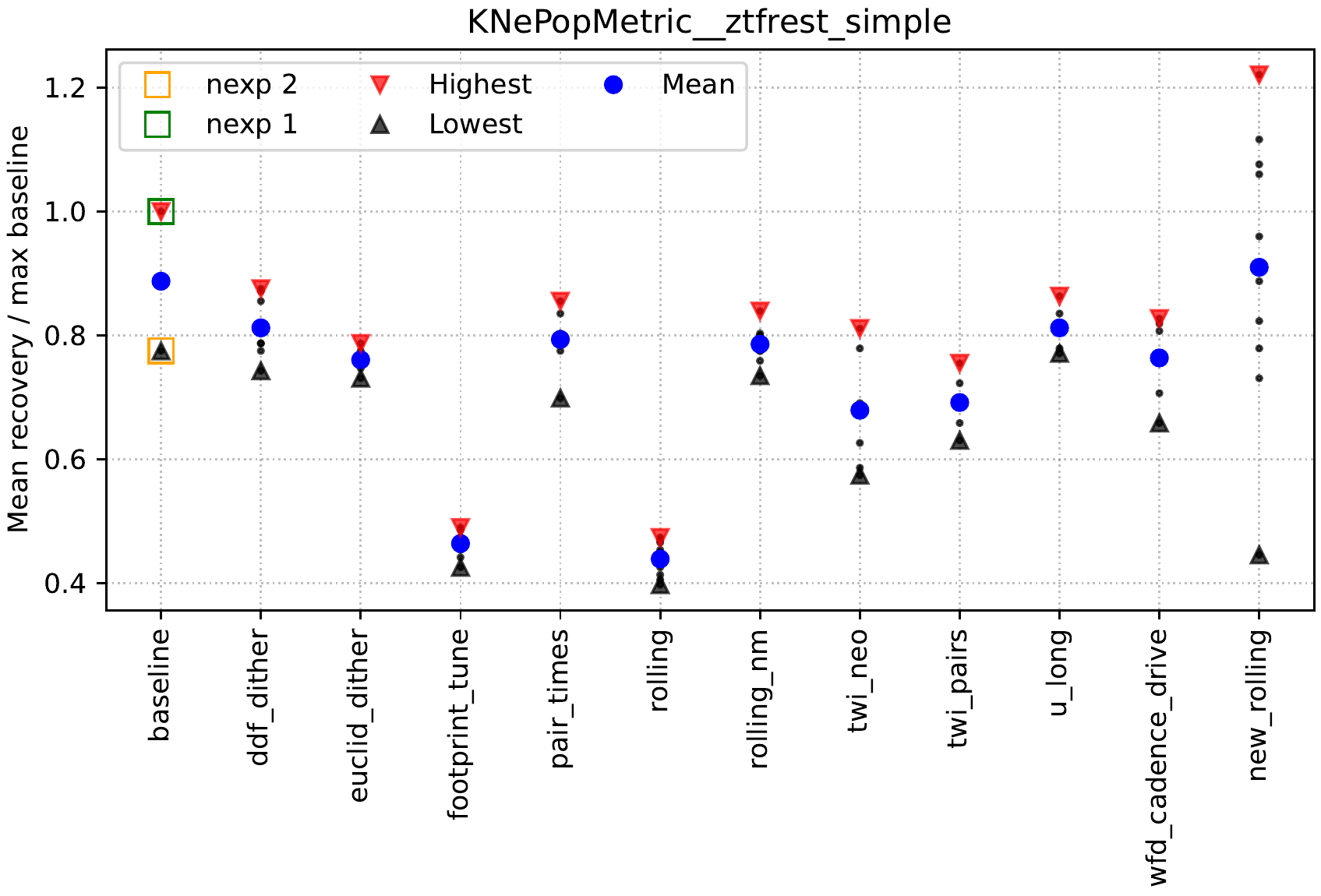}
 \includegraphics[width=0.49\textwidth]{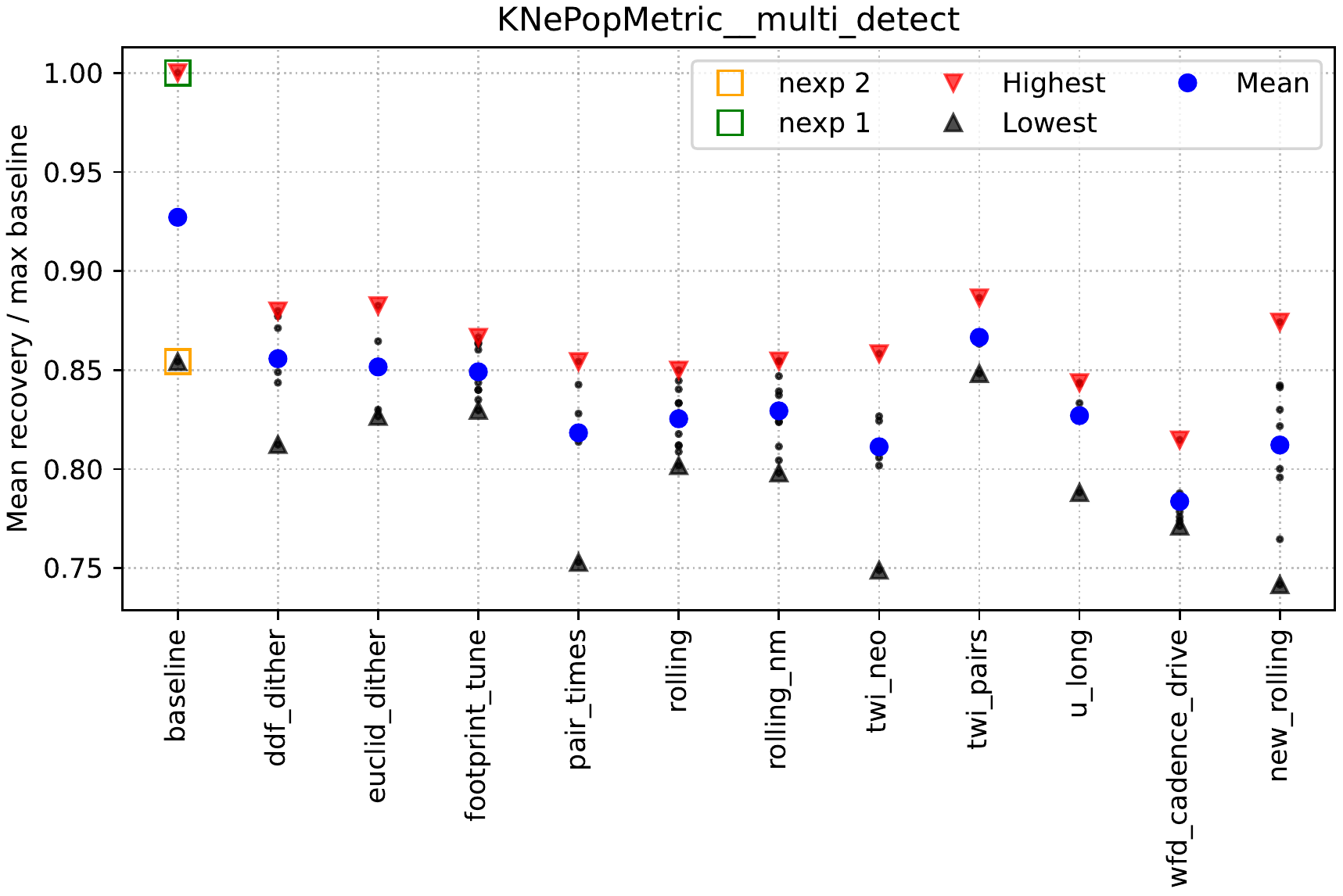}
 \includegraphics[width=0.49\textwidth]{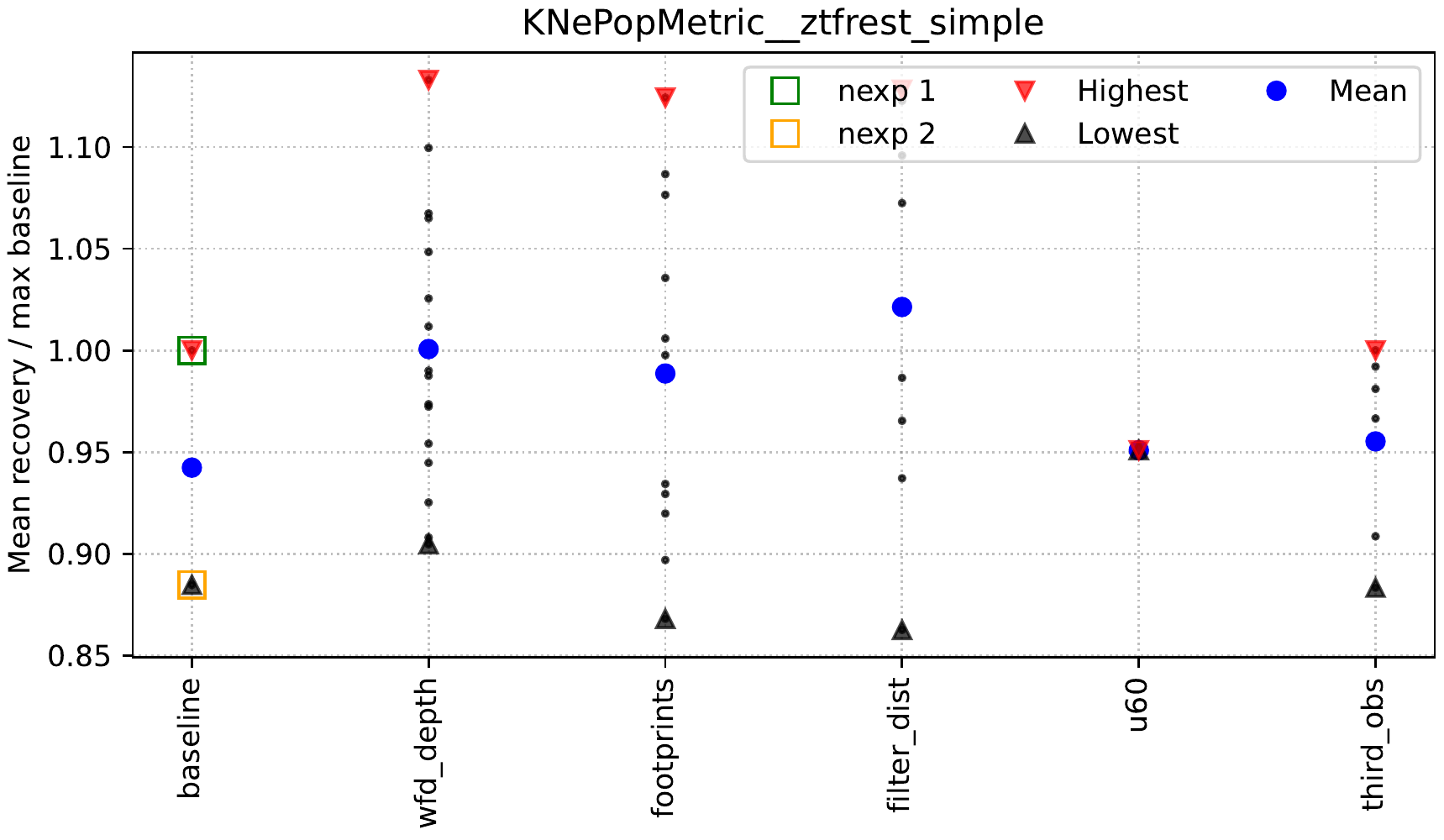}
 \includegraphics[width=0.49\textwidth]{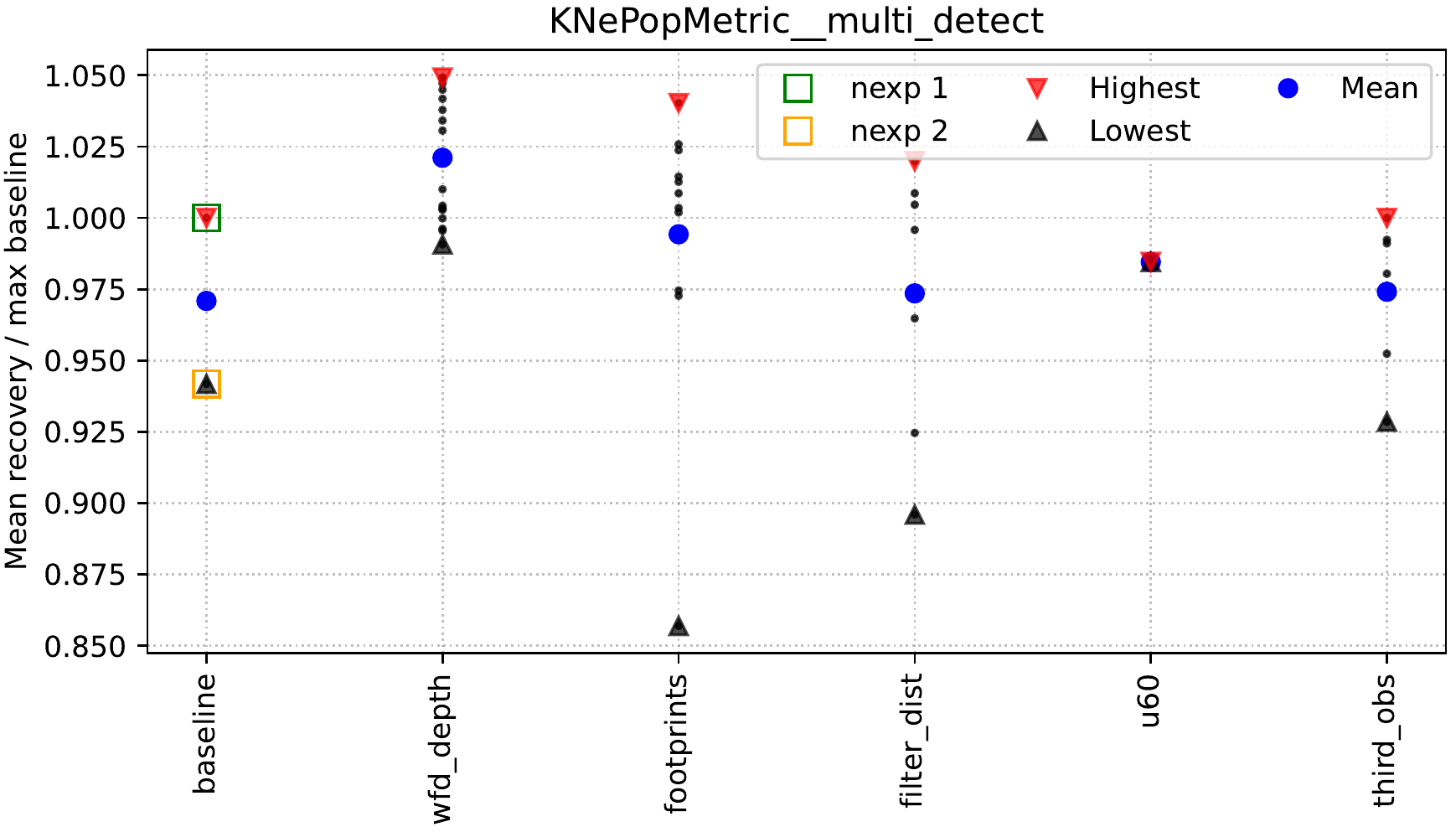}  
  \caption{For each cadence family, we report the ratio between the recovery fraction of the individual cadences and the maximum recovery fraction from the best baseline cadence (baseline\_nexp1). Blue dots indicate the mean of the cadences' recovery fractions, red triangles their maximum, and black triangles their minimum. The two-snaps baseline cadence (baseline\_nexp2) performs systematically worse than the single-exposure baseline cadence. Simulations part of FBS v1.7 (top) and of FBS v1.5 (bottom) are considered separately.}
 \label{fig:metric_results}
\end{center}
\end{figure*}

\subsection{Kilonova models}
\label{subsec:models}

In this work, we considered kilonova models from the grid generated with the three-dimensional radiation transfer simulation code \texttt{POSSIS} \citep{Bulla2019}. The model grid allowed us to explore a diversity of intrinsic properties, such as ejecta masses, as well as different viewing angles to the system. 

First, we injected synthetic light curves using a single model: the GW170817-like kilonova (dynamical ejecta mass $M_{\text{dyn}} = 0.005 M_{\odot}$, disk wind mass $M_{\text{wind}} = 0.050 M_{\odot}$, and viewing angle $\iota = 25.8{\degree}$). A half-opening angle $\phi = 30\degree$ for the lanthanide-rich region is used for this model and all the other models considered in this work. Second, we injected a population of kilonovae with the same ejecta masses of the GW170817-like model but viewed from eleven viewing angles, uniformly distributed in $cos(\iota)$. Third, we explored kilonova detectability in an optimistic and a pessimistic scenarios, in which the kilonova properties make it particularly bright or dim, respectively. Ejecta masses were chosen to be physically realistic as determined by numerical relativity simulations, with $M_{\text{dyn}} = 0.020 M_{\odot}$, $M_{\text{wind}} = 0.090 M_{\odot}$ for the optimistic scenario and $M_{\text{dyn}} = 0.005 M_{\odot}$, $M_{\text{wind}} = 0.010 M_{\odot}$ for the pessimistic scenario.



\section{Results}
\label{sec:results}

\subsection{GW170817-like kilonova light curves}

Cadences were made available in several releases and were grouped into ``families", in which ideas that deviate from the baseline cadence were implemented and encompass parameter variations, for example in the area of the footprint. Detailed information about simulations can be found in official Rubin Observatory online resources\footnote{for example \url{https://github.com/lsst-pst/survey_strategy}}. 
Fig.\,\ref{fig:metric_results} shows results obtained by injecting $5 \times 10^5$ synthetic GW170817-like kilonova light curves, uniformly distributed in volume out to a luminosity distance of 1.5\,Gpc, in \texttt{OpSim} simulated cadences part of the v1.5\footnote{\url{https://github.com/lsst-sims/sims_featureScheduler_runs1.5}} (bottom panels) and v1.7\footnote{\url{https://github.com/lsst-sims/sims_featureScheduler_runs1.7}} and v1.7.1\footnote{\url{https://community.lsst.org/t/survey-simulations-v1-7-1-release-april-2021/4865}} releases (upper panels). The results displayed in Fig.\,\ref{fig:metric_results} were obtained using the \texttt{multi\_detect} and the \texttt{ztfrest\_simple} metrics described in \S\ref{subsec:metrics}.

In all simulations,  the best baseline cadence entails individual 30\,s exposures (baseline\_nexp1). 
The baseline cadence where $2\times 15$\,s snaps (baseline\_nexp2) performs consistently worse.   
The number of recovered kilonovae in cadence families simulated as part of the v1.5 cadence release (bottom panels) are relatively similar, with results comparable with the best baseline cadence within 15\%. When looking at v1.7 cadences, it is evident that the best baseline performs distinctly better than any other cadence in terms of kilonova detection (\texttt{multi\_detect} metric; top-right panel of Fig.\,\ref{fig:metric_results}).  The baseline cadence does a better job than most cadence families, also according to the \texttt{ztfrest\_simple} metric. We found that rolling cadences, in which a smaller fraction of the footprint is observed in each season at higher cadence, perform significantly ($\sim 50-60\%$) worse as coded for the v1.7 release than the baseline cadence\footnote{Significant changes to the \texttt{Opsim} approach to simulate rolling cadence strategies were implemented for v1.7, such that these simulations should be considered more reliable (see Bianco et al.; front paper of this Focus Issue)}. However, rolling cadences part of the v1.7.1 release, indicated as ``new\_rolling'' in the figure,  perform up to $\sim 20\%$ better than the baseline cadence (in the Figure, uncertainties are in the order of 5-10\%).
In order to compare baseline and rolling cadences with a higher statistical significance, we ran simulations in which the number of injected sources was increased to $10^6$, using a variety of surrogate models. A summary of the results can be found in Tab.\,\ref{tab:bas_vs_roll}. 

\begin{figure}[t]
\begin{center}
 \includegraphics[width=\columnwidth]{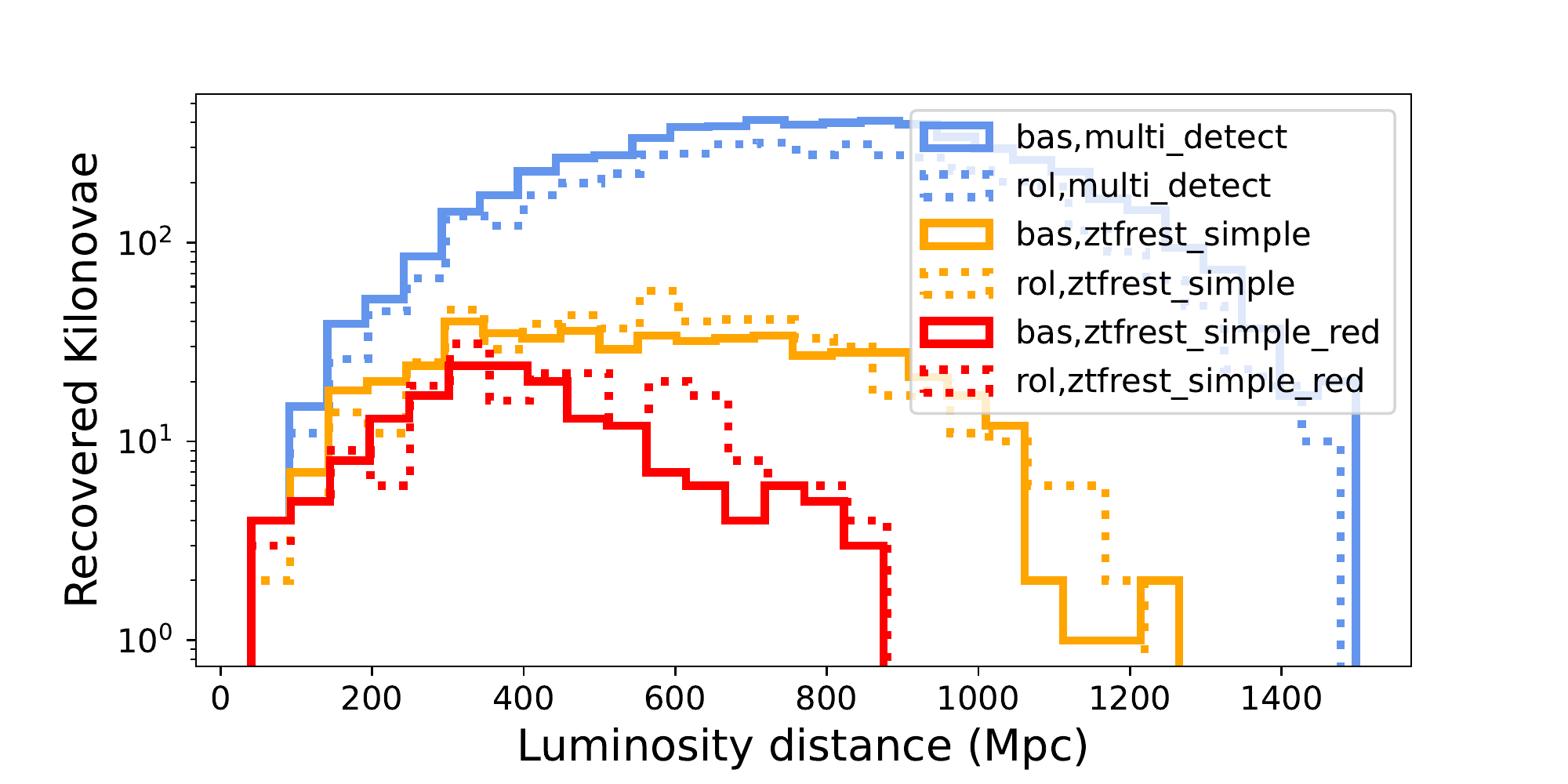}
  \includegraphics[width=\columnwidth]{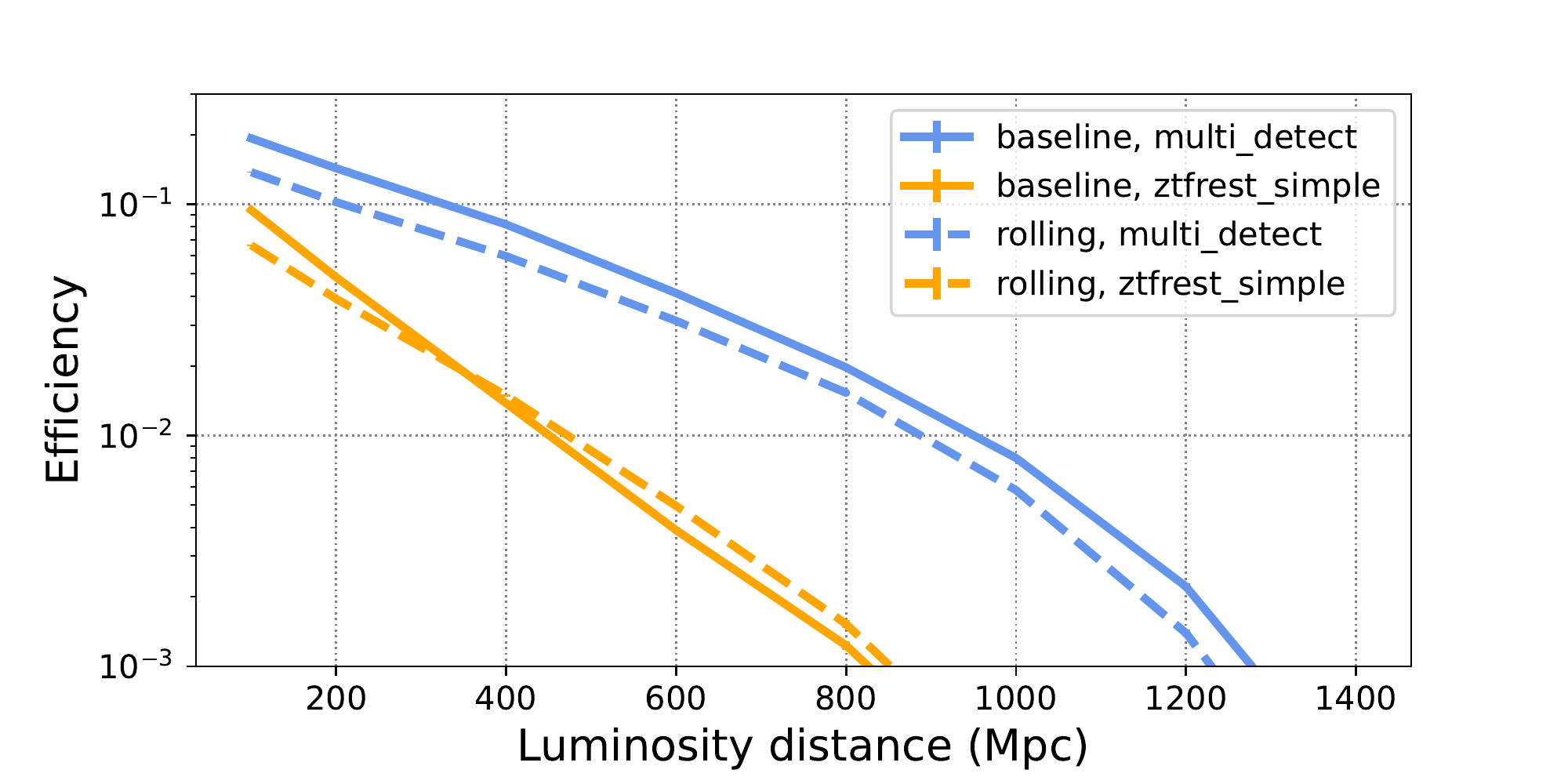}
  \includegraphics[width=\columnwidth]{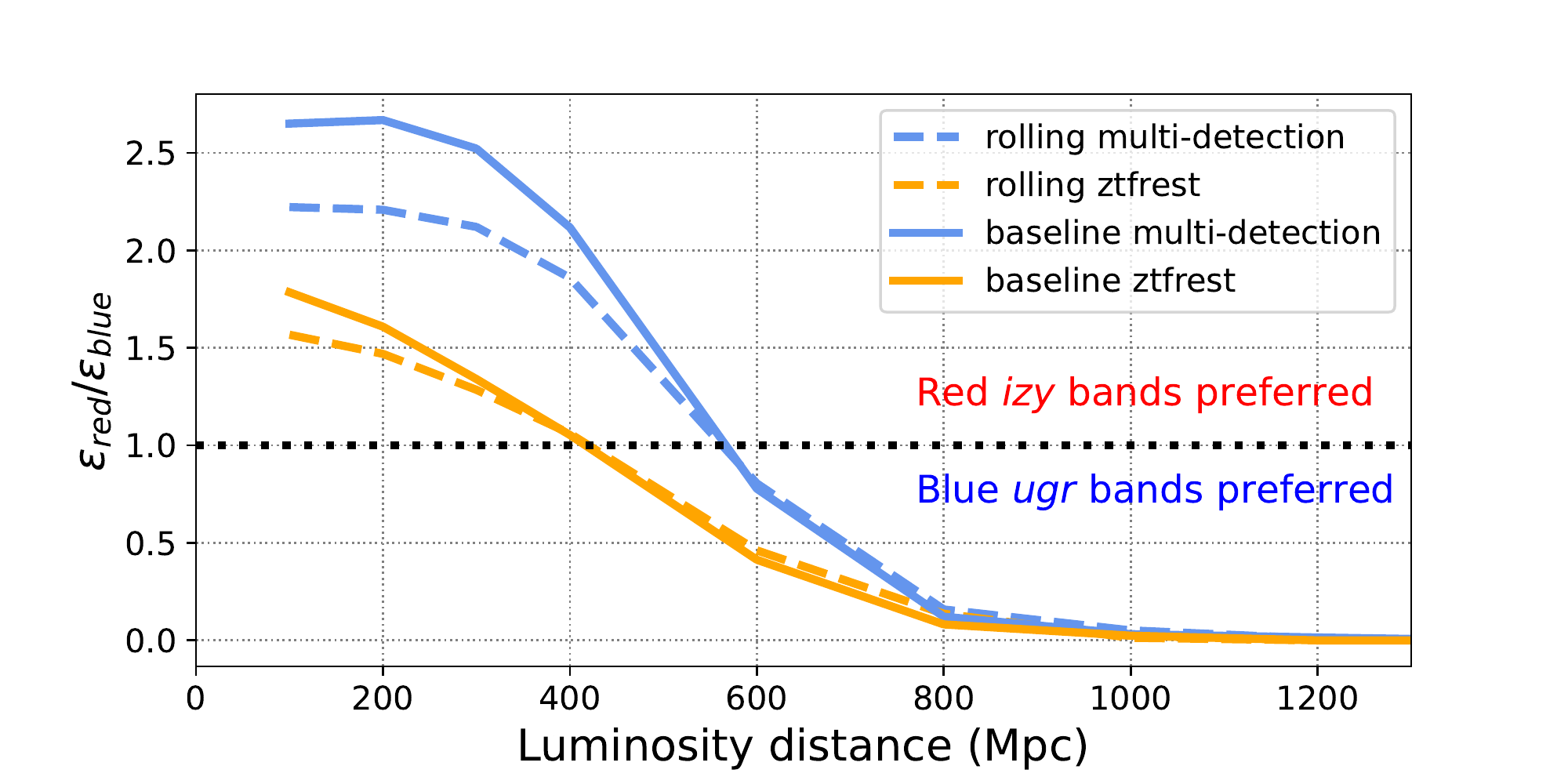}
  \caption{{\it Top:} Distribution of recovered kilonovae using a simple multi-detection metric (cyan), a \texttt{ZTFReST}-like metric (orange), and a \texttt{ZTFReST}-like metric applied only to red ($izy$) bands (red lines). One million sources were injected uniformly distributed in volume between 10\,Mpc and 1.5\,Gpc. {\it Center:} Efficiency as a function of luminosity distance; $5\times 10^5$ sources were injected at regular intervals. For the \texttt{ztfrest\_simple} metric, due to the rapidly growing rate at the edge of the sensitive volume, small differences between detection efficiencies for the rolling and the baseline cadence beyond $\sim 400$\,Mpc is enough to yield an improvement of $\sim 20\%$ in kilonova detection. 
  {\it Bottom:} Ratio between the detection efficiency in redder $izy$ bands and in bluer $ugr$ bands for multi-detection and \texttt{ZTFReST}-like metrics. Employing redder filters presents clear advantages at lower distances, where spectroscopic and multi-wavelength follow-up observations are possible.
  }
 \label{fig:distribution}
\end{center}
\end{figure}

When we injected GW170817-like kilonovae,  the baseline cadence performed better than the best rolling cadence\footnote{six\_stripe\_scale0.90\_nslice6\_fpw0.9\_nrw0.0v1.7\_10yrs} at any distance with the \texttt{multi\_detect} metric (Fig.\,\ref{fig:distribution}, central panel), but the rolling cadence outperforms the baseline cadence beyond $\sim 400$\,Mpc with the \texttt{ztfrest\_simple} metric. This means that a rolling cadence could enable the identification of a few more kilonova candidates than the baseline cadence at large distances. In total, 32--334 kilonovae can be expected to be detectable with the baseline cadence and 23--238 with the rolling cadence, assuming that all kilonovae are similar to the observed GW170817. However, the number of kilonovae recognizable to be fast transients (\texttt{ztfrest\_simple} metric) would be 3--29 and 3--32 for the baseline and rolling cadences, respectively.

Nearby kilonovae have the potential of being recognized sooner, associated with hosts at known redshifts, and can be better characterized with follow-up observations than distant (fainter) sources. To better explore detectability of nearby kilonovae, we injected $10^6$ GW170817-like kilonovae uniformly distributed in volume out to 300\,Mpc, which is within the distance range where the best Wide Fast Deep (WFD) baseline cadence performs better than any of the rolling cadences simulated so far. With the best baseline cadence, we can expect up to 101 kilonovae to be detectable at least twice in this distance range and up to 31 could be recognized to be fast-fading in at least one band. In the simulations, about $68\%$ of kilonovae were found to be fast-fading in red $izy$ bands (\texttt{ztfrest\_simple\_red} metric) and 44\% in blue $ugr$ bands (\texttt{ztfrest\_simple\_blue} metric). Only 37\% of kilonovae found in $izy$ bands were detected at least 4 times in $ugr$ bands (Fig.\,\ref{fig:distribution}, bottom panel).  The combination of transient detection, color information, and possible association with a catalogued nearby galaxy (which enables an estimate of the transient's absolute magnitude, expected to be fainter than $M \sim -16$ for most kilonovae) can lead to the identification of solid kilonova candidates. For the fraction of events that are relatively nearby (below 300\,Mpc), they can be followed up spectroscopically with $\gtrsim$\,8-m class telescopes such as the Very Large Telescope (VLT), Gemini, Keck, or with the upcoming SoXS at ESO New Technology Telescope (NTT), which was designed specifically for LSST transient classification, to be classified \citep{2016SPIE.9908E..41S}. In summary, our analysis suggests that employing more observations in redder bands is preferred to maximize scientific return.

\subsection{Exploring the kilonova light curve parameter space}

Multi-messenger observations of GW170817 constrained the viewing angle to be $\iota = 32_{-13}^{+10}$\,deg \citep{Finstad2018}, see also \citet{2020ApJ...888...67D}. Superluminal motion from radio observations suggests a lower value for the viewing angle of about 15--20\,deg \citep{Mooley2018Nat, Ghirlanda2019Sci}.

However, merging BNS systems can be oriented in any direction with respect to the observer. We compared the performance of baseline and rolling cadences also by injecting synthetic light curves, from the grid of kilonova models simulated with \texttt{POSSIS}, with the same intrinsic parameters as the GW170817-like model (\S\ref{subsec:models}), but at different viewing angles. According to our simulations, up to 15 (17) kilonovae should be identified as fast transients in the baseline (rolling) cadence, while up to 176 (127) kilonovae should be detectable at least twice.

Finally, we assessed kilonova detectability for ``pessimistic" and ``optimistic'' kilonova models, in which the ejecta masses make the optical emission particularly faint or bright (see \S\ref{subsec:models}). For the pessimistic case, only a handful of kilonovae are expected to be present in Rubin images, with at most 5 kilonovae expected to be recognizable as fast transients. On the other hand, the optimistic scenario could result in $>50$ kilonovae found to evolve rapidly with the baseline cadence and $>60$ with the currently best rolling cadence. A better understanding of the kilonova luminosity function is required to set more precise serendipidous kilonova discovery expectations.

\begin{table*}[h]
    \centering
\caption{Kilonova recovery efficiencies ($\epsilon$), calculated with a number of metrics, for the best baseline cadence 
and the best rolling cadence. 
The efficiency was then converted into the number of expected kilonovae using the BNS merger rate $R=320_{-240}^{+490}$\,Gpc$^{-3}$\,yr$^{-1}$ from the GWTC-2 catalog \citep{LIGO_GWTC2_pop_2020arXiv}, where N$_{\text{KN}}$ corresponds to the median rate and N$_{\text{KN,min}}$, N$_{\text{KN,max}}$ correspond to the 90\% symmetric credible intervals, taking the uncertainty in $\epsilon$ into account. A duration of 10\,yr for the WFD survey was assumed. }
    \label{tab:bas_vs_roll}
    \begin{tabular}{llcccccccc}
    \hline\hline
     Kilonova & Metric & $\epsilon_{\text{baseline}}$ & $\epsilon_{\text{rolling}}$ & N$_{\text{KN}}$ & N$_{\text{KN}}$ & N$_{\text{KN,min}}$ & N$_{\text{KN,min}}$ & N$_{\text{KN,max}}$ & N$_{\text{KN,max}}$ \\
     model & & $\times 10^4$ & $\times 10^4$ & baseline & rolling & baseline & rolling & baseline & rolling\\
    \hline
    GW170817 & multi\_detect & $60.5 \pm 0.8$ & $43.1 \pm 0.7$ & 130 & 93 & 32 & 23 & 334 & 238  \\
$M_{\text{dyn}} = 0.005 M_{\odot}$ & blue\_color\_detect & $6.8 \pm 0.3$ & $6.4 \pm 0.2$ & 15 & 14 & 4 & 3 & 38 & 36  \\
$M_{\text{wind}} = 0.050 M_{\odot}$ & red\_color\_detect & $2.9 \pm 0.2$ & $3.1 \pm 0.2$ & 6 & 7 & 1 & 2 & 17 & 18  \\
 & ztfrest\_simple & $5.2 \pm 0.2$ & $5.6 \pm 0.2$ & 11 & 12 & 3 & 3 & 29 & 32  \\
 & ztfrest\_simple\_blue & $3.8 \pm 0.2$ & $4.4 \pm 0.2$ & 8 & 9 & 2 & 2 & 22 & 25  \\
 & ztfrest\_simple\_red & $1.7 \pm 0.1$ & $2.1 \pm 0.1$ & 4 & 4 & 1 & 1 & 10 & 12  \\
    \hline
GW170817 & multi\_detect & $1306.3 \pm 3.6$ & $931.2 \pm 3.0$ & 40 & 28 & 10 & 7 & 101 & 72  \\
$<300$\,Mpc & blue\_color\_detect & $141.0 \pm 1.2$ & $143.8 \pm 1.2$ & 4 & 4 & 1 & 1 & 11 & 11  \\
 & red\_color\_detect & $369.5 \pm 1.9$ & $312.0 \pm 1.8$ & 11 & 10 & 3 & 2 & 29 & 24  \\
 & ztfrest\_simple & $400.3 \pm 2.0$ & $334.1 \pm 1.8$ & 12 & 10 & 3 & 3 & 31 & 26  \\
 & ztfrest\_simple\_blue & $176.7 \pm 1.3$ & $173.0 \pm 1.3$ & 5 & 5 & 1 & 1 & 14 & 13  \\
 & ztfrest\_simple\_red & $272.9 \pm 1.6$ & $244.5 \pm 1.6$ & 8 & 7 & 2 & 2 & 21 & 19  \\
 \hline
    GW170817  & multi\_detect & $31.8 \pm 0.6$ & $22.7 \pm 0.5$ & 68 & 49 & 17 & 12 & 176 & 127  \\
    viewing & blue\_color\_detect & $3.4 \pm 0.2$ & $3.4 \pm 0.2$ & 7 & 7 & 2 & 2 & 20 & 19  \\
 angles & red\_color\_detect & $1.4 \pm 0.1$ & $1.7 \pm 0.1$ & 3 & 4 & 1 & 1 & 8 & 10  \\
 & ztfrest\_simple & $2.6 \pm 0.2$ & $2.9 \pm 0.2$ & 6 & 6 & 1 & 1 & 15 & 17  \\
 & ztfrest\_simple\_blue & $1.8 \pm 0.1$ & $2.3 \pm 0.1$ & 4 & 5 & 1 & 1 & 10 & 13  \\
 & ztfrest\_simple\_red & $1.0 \pm 0.1$ & $1.1 \pm 0.1$ & 2 & 2 & 0 & 1 & 6 & 7  \\
 \hline
Pessimistic  & multi\_detect & $8.9 \pm 0.3$ & $6.5 \pm 0.2$ & 19 & 14 & 5 & 3 & 50 & 37  \\
$M_{\text{dyn}} = 0.005 M_{\odot}$ & blue\_color\_detect & $0.8 \pm 0.1$ & $0.9 \pm 0.1$ & 2 & 2 & 0 & 0 & 5 & 5  \\
$M_{\text{wind}} = 0.010 M_{\odot}$  & red\_color\_detect & $0.3 \pm 0.1$ & $0.4 \pm 0.1$ & 1 & 1 & 0 & 0 & 2 & 3  \\
 & ztfrest\_simple & $0.5 \pm 0.1$ & $0.6 \pm 0.1$ & 1 & 1 & 0 & 0 & 3 & 4  \\
 & ztfrest\_simple\_blue & $0.4 \pm 0.1$ & $0.4 \pm 0.1$ & 1 & 1 & 0 & 0 & 2 & 2  \\
 & ztfrest\_simple\_red & $0.2 \pm 0.1$ & $0.3 \pm 0.1$ & 0 & 1 & 0 & 0 & 1 & 2  \\

\hline
Optimistic  & multi\_detect & $116.6 \pm 1.1$ & $87.0 \pm 0.9$ & 251 & 187 & 62 & 46 & 641 & 479  \\
$M_{\text{dyn}} = 0.020 M_{\odot}$ & blue\_color\_detect & $11.9 \pm 0.3$ & $12.9 \pm 0.4$ & 26 & 28 & 6 & 7 & 67 & 72  \\
$M_{\text{wind}} = 0.090 M_{\odot}$ & red\_color\_detect & $5.2 \pm 0.2$ & $5.9 \pm 0.2$ & 11 & 13 & 3 & 3 & 29 & 34  \\
 & ztfrest\_simple & $9.2 \pm 0.3$ & $10.8 \pm 0.3$ & 20 & 23 & 5 & 6 & 52 & 61  \\
 & ztfrest\_simple\_blue & $6.7 \pm 0.3$ & $8.3 \pm 0.3$ & 14 & 18 & 3 & 4 & 38 & 47  \\
 & ztfrest\_simple\_red & $3.2 \pm 0.2$ & $4.2 \pm 0.2$ & 7 & 9 & 2 & 2 & 18 & 24  \\
 \hline
    \end{tabular}
\end{table*}

\section{Conclusion}
\label{sec:conclusion}

Rubin Observatory has a great potential of revealing a population of kilonovae during the WFD survey, in addition to discoveries made following up GW triggers. We injected synthetic kilonova light curves into simulated Rubin observations to assess which ones of the available cadences can maximize serendipitous kilonova discovery. We demonstrated that, for the WFD survey, the simulated baseline cadence with single 30\,s exposures should be greatly preferred over $2 \times 15$s consecutive snaps for kilonova discovery.

Rolling cadences are expected to be particularly suitable for fast transient discovery \citep[e.g.,][]{Andreoni2019PASP}. We found that the development of rolling cadences has significantly improved from \texttt{OpSim} version v1.7 to v1.7.1. While this indicates progress, the baseline plan may still be preferred over any other cadence family currently available due to a larger efficiency at detecting more nearby (and therefore brighter) fast transients, easier to follow-up and classify with other telescopes. We recommend simulating new rolling cadences further optimizing the algorithms used in v1.7.1, possibly maximizing the exposure time in each band (barring $u$-band) rather than using snap pairs. 

We found strong evidence that red $izy$ bands are preferred for kilonova discovery at distances below 300\,Mpc, in agreement with the results of other studies such as, for example, \cite{Almualla2021} and \cite{Sagues2021}. This is expected because kilonovae appear as red and rapidly-reddening transients due to heavy $r$-process elements synthesised in neutron-rich ejecta. At redder wavelengths, kilonova light curves can be brighter and longer-lived, especially if the system is viewed from equatorial viewing angles \citep[e.g.,][]{Bulla2019}. Very rapid ``blue" kilonovae could be found at larger distances (Fig.\,\ref{fig:distribution}) due to the greater sensitivity of $g$ and $r$ filters, however, these kilonovae might be rarer and more difficult to classify spectroscopically.
Therefore, we recommend that the number of $izy$ observations is increased in the WFD cadence plan. Such red-band observations would be particularly effective, scientifically, if coupled with at least one observation in $g$- (preferred) or $r$-band on the same night, so that kilonovae can be separated photometrically from other transients and their temperature evolution can be measured. A recommendation for same-night multi-band photometry in LSST has already been put forward for example by \cite{Andreoni2019PASP} and \cite{Bianco2019}. In particular, \cite{Bianco2019} address the advantages of acquiring sets of three exposures per field in the same night, in two filters and appropriately spaced in time, towards rapid identification of rare fast transients.

Major uncertainty in the results of this work results from our limited understanding of the BNS merger rate and the kilonova luminosity function. Systematic kilonova searches during gravitational-wave follow-up \cite[e.g.,][]{Kasliwal2020}, short GRB follow-up \cite[e.g.,][]{Gompertz2018, Rossi2020}, and un-triggered wide-field surveys \citep[e.g.,][]{Doctor2017, AnKo2020, AnCo2021arXiv, McBrien2021MNRAS} are expected to improve those measurements significantly before Rubin Observatory's first light.

\section*{Acknowledgments}

We thank Peter Yoachim and Lynne Jones. This paper was created in the nursery of the Rubin LSST  Transient and Variable Star Science Collaboration \footnote{\url{https://lsst-tvssc.github.io/}}. The authors acknowledge the support of the Vera C. Rubin Legacy Survey of Space and Time Transient and Variable Stars Science Collaboration that provided opportunities for collaboration and exchange of ideas and knowledge and of Rubin Observatory in the creation and implementation of this work.
The authors acknowledge the support of the LSST Corporation, which enabled the organization of many workshops and hackathons throughout the cadence optimization process by directing private funding to these activities.

\noindent M.~W.~C acknowledges support from the National Science Foundation with grant number PHY-2010970.
M. B. acknowledges support from the Swedish Research Council (Reg. no. 2020-03330).
A.G, A.S.C and E. C. K. acknowledge support from the G.R.E.A.T. research environment funded by {\em Vetenskapsr\aa det}, the Swedish Research Council, under project number 2016-06012, and support from The Wenner-Gren Foundations.

\noindent This research uses services or data provided by the Astro Data Lab at NSF's National Optical-Infrared Astronomy Research Laboratory. NOIRLab is operated by the Association of Universities for Research in Astronomy (AURA), Inc. under a cooperative agreement with the National Science Foundation.

\software{LSST metrics analysis framework \citep[MAF;][]{Jones2014SPIE}; Astropy \citep{2013A&A...558A..33A}; JupyterHub\footnote{\url{https://jupyterhub.readthedocs.io/en/stable/index.html}}}

\bibliographystyle{aasjournal}
\begin{small}
\bibliography{references}
\end{small}

\end{document}